\documentclass[]{svjour2}                    
\usepackage{graphicx}
\usepackage{amsmath}
\usepackage{amssymb}
\usepackage{bbm}

\begin{document}

\title{Evolutionary accessibility of modular fitness landscapes
}


\author{Benjamin Schmiegelt \and Joachim Krug}


\institute{B. Schmiegelt \and J. Krug \at
              Institute for Theoretical Physics, University of
              Cologne, K\"oln, Germany \\
              Tel.: +49 221 2818\\
	      Fax: +49 221 5159\\
              \email{schmiegb@smail.uni-koeln.de, krug@thp.uni-koeln.de}          
}

\date{Received: date / Accepted: date}

\maketitle

\begin{abstract}
A fitness landscape is a mapping from the space of genetic sequences, which is modeled here as a 
binary hypercube of dimension $L$, to the real numbers. We consider random models of fitness landscapes,
where fitness values are assigned according to some probabilistic rule, and study the statistical
properties of pathways to the global fitness maximum along which fitness increases monotonically.
Such paths are important for evolution because they are the only ones that are accessible to an adapting
population when mutations occur at a low rate. The focus of this work is on the block model introduced
by A.S. Perelson and C.A. Macken [Proc. Natl. Acad. Sci. USA 92:9657 (1995)] where the 
genome is decomposed into disjoint sets of loci (`modules') 
that contribute independently to fitness, and fitness
values within blocks are assigned at random. We show that the number of accessible paths can be 
written as a product of the path numbers within the blocks, which provides a detailed analytic
description of the path statistics. The block model can be viewed as a special case of Kauffman's
NK-model, and we compare the analytic results to simulations of the NK-model with different genetic
architectures. We find that the mean number of accessible paths in the different versions of the model
are quite similar, but the distribution of the path number is qualitatively different in the block model
due to its multiplicative structure. A similar statement applies to the number of local fitness maxima
in the NK-models, which has been studied extensively in previous works. The overall evolutionary
accessibility of the landscape, as quantified by the probability to find at least one
accessible path to the global maximum, is dramatically lowered by the modular structure.
\end{abstract}

\section{Introduction}

Random mutations on different scales of the genome introduce non-deter-ministic genetic diversity to an evolving population, opening up new pathways for exploration of the genotypic space. At the same time selection restricts the number of possible evolutionary trajectories in a deterministic manner. From the interplay between these two contrary forces
arises the question whether evolution as a whole is predictable and reproducible
\cite{Travisano1995,Hall2002,Jain2007,ConwayMorris2010,Lobkovsky2012,Szendro2013}.

In an environment of strong selective pressure and weak mutation rates and/or small population size, possible steps towards higher fitness are largely limited by the structure of the fitness landscape on which adaptation takes place. In this \textit{strong selection weak mutation (SSWM)} regime populations cannot overcome fitness valleys by generating multiple mutants. Rather, each single mutation, introduced one at a time, has to prove beneficial, resulting in an uphill walk on the fitness landscape \cite{Gillespie1983,Macken1989,Macken1991,Flyvbjerg1992,Orr2002,Neidhart2011}.

On a fully additive landscape, where each genetic locus contributes independently to the overall fitness, beneficial mutations can occur in any order, which implies many possible mutational pathways.
However, often the fitness contributions of different loci are not independent. Mutations whose 
effect depends on the state of other loci (the \textit{genetic background}) are known as epistatic \cite{Phillips2008}. Cases in which not only the value of fitness change but also the sign of change (beneficial or deleterious) depends on the state of other loci are known as sign-epistatic
\cite{Weinreich2005,Poelwijk2007,Kvitek2011}. Landscapes with sign-epistatic interactions tend to be rugged and may have multiple local optima \cite{Poelwijk2010,Crona2013}. Recent empirical evidence
suggests that sign epistasis is common in biological entities ranging from single proteins
\cite{Weinreich2006} to entire organisms \cite{Franke2011}, see \cite{Szendro2013a} for review. 

As part of the general problem of understanding possible evolutionary outcomes and pathways, we here focus on the question: How does epistasis influence the accessibility of the global fitness 
maximum in the SSWM regime? In recent work, this question has been addressed for several well 
known models of fitness landscapes \cite{Kloezer2008,Carneiro2010,Franke2011,Franke2012,Hegarty2012,Nowak2013,Berestycki2013,Roberts2013}, 
in particular the House-of-Cards/Random Energy model
\cite{Kingman1978,Kauffman1987}, the Rough Mt. Fuji model
\cite{Aita2000,Franke2010} and Kauffman's NK-model \cite{Kauffman1989,Kauffman1993}. 
In the NK-model each genetic locus interacts with a neighborhood of $k$ other loci, and different
genetic architectures can be realized through different ways of chosing the neighbors.  
Despite its simplicity and lack of biological detail the NK-model has proven to be useful for parametrizing
empirical fitness landscapes, thus providing a quantitative characterization of the strength
and type of epistatic interactions in these data sets \cite{Franke2011}. 
The versatility of the model can be further increased by considering linear superpositions of NK-landscapes
with different values of $k$ \cite{Neidhart2013}. 

Here we will focus on fitness landscapes that have a modular structure, in that 
the genetic loci are divided into disjoint sets, called blocks, which contribute independently to the overall fitness. Such a model was first introduced by Perelson and Macken \cite{Perelson1995},
and it can be viewed as a special case of Kauffman's NK-model. We will see that the block structure
significantly facilitates analytic calculations, to the extent that a detailed characterization
of the full probability distribution of the number of accessible mutational pathways becomes
possible. Surprisingly, the exact expression for the mean number of accessible paths,
similar to the mean number of optima derived in \cite{Perelson1995}, turns out to closely match
the numerical estimates obtained for other versions of the NK-model \cite{Franke2011,Franke2012}.
At the same time the fluctuations in these quantities show a strong dependence on the genetic
architecture, leading in particular to a very low evolutionary accessibility of the block model
landscape compared to the NK-model with random (non-modular) interactions studied previously \cite{Franke2012}.

In the next section we explain the basic mathematical concepts required for the description
of genotype spaces and fitness landscapes, and introduce the models of interest. Our results
on the evolutionary accessibility of modular landscapes are presented in Sect.~\ref{Results}, 
and the paper concludes with a summary and an outlook in Sect.~\ref{Summary}.

\section{Fitness landscapes and their maxima}

In the SSWM regime the genetic variability in a population is small and it can be assumed that all individuals have the same genotype most of the time, apart from the transient appearance of single
new mutations. The genotype of a population can be modeled as a binary sequence of length $L$, $\sigma = (\sigma_1,\ldots,\sigma_L)$ where each $\sigma_i$ is either $1$ or $0$ representing two different alleles at locus $i$ or a wild type and a mutated type. The space of all possible genotypes 
is then the binary hypercube $\mathbb H_2^L = \{0,1\}^L$, which we extend into a normed space by introducing the Hamming norm $\Vert\sigma\Vert = \sum_{i=1}^L \sigma_i$ and the induced Hamming metric $d(\sigma,\theta) = \Vert\sigma-\theta\Vert = \sum_{i=1}^L \vert \sigma_i-\theta_i \vert$. This metric represents the number of loci in which two genotypes differ and hence the minimal number of point mutations needed to reach one from the other. For future reference we define the \textit{antipodal} or reversal sequence
$\bar \sigma$ of a genotype $\sigma$ through $\bar{\sigma_i} = 1 - \sigma_i$. A genotype and its  antipodal sequence are maximally distant from each other, $d(\sigma,\bar{\sigma}) = L$ for all $\sigma$.

Since we only consider point mutations, we define the mutation operator $\Delta_i$ which mutates locus $i$ as
\begin{equation}
\label{mutations}
\Delta_i \sigma := (\sigma_1,\ldots,\sigma_{i-1},1-\sigma_i,\sigma_{i+1},\ldots,\sigma_L). 
\end{equation}
We can extend this notion to simultaneous mutations at several
loci. Let $M = \{M_1,\ldots,M_m\} \subseteq \{1,\ldots,L\}$ be the
set of loci that are to be mutated. We then denote the group mutation operator as $\Delta_M\sigma := \Delta_{M_1}\ldots\Delta_{M_m}\sigma$. 

A fitness landscape on the space of sequences of length $L$ is a mapping from $\mathbb H_2^L$ into the real numbers $F:\; \mathbb H_2^L \rightarrow \mathbb{R}$. We use the notation $\Delta_M F(\sigma) := F(\Delta_M\sigma) - F(\sigma)$ to refer to the change in fitness by mutating all loci in $M$ starting from genotype $\sigma$. By applying each single locus mutation to each genotype on the fitness landscape we generate an $L$-dimensional real vector field $\Delta F$ on the genotype space, $\Delta F(\sigma) = (\Delta_1 F(\sigma), \ldots, \Delta_L F(\sigma))$. This field determines the effect of every possible mutation at each point of the fitness landscape. It defines the fitness landscape uniquely up to a constant. Therefore all relevant properties of the fitness landscape are determined by $\Delta F$. However not all mappings $\Delta F: \mathbb H_2^L \rightarrow \mathbb{R}^L$ are valid mutation fields of a fitness landscape.

In the following we introduce the fitness landscape models of 
interest in this work. They are random field models in the sense of \cite{Stadler1999} and 
bear a close resemblance to spin glass models of statistical physics \cite{Mezard1987,Bovier2006}. 
A common way of quantifying the ruggedness of such fitness or energy landscapes is through the
number of local maxima, and we compile some known results for this quantity for the different models
below.

\subsection{House-of-Cards model}

In the House-of-Cards (HoC) model every fitness value is drawn identically and independently from a real-valued probability distribution \cite{Kingman1978,Kauffman1987,Kauffman1993}.
Since only the sign of fitness change is relevant to accessibility, it is sufficient to consider the HoC model as a random rank order on the genotype space. The properties discussed here 
therefore do not depend on the chosen probability distribution. Up to a change of sign  
the HoC landscape is equivalent\footnote{For further discussion of the relation between fitness landscapes and spin glass 
models we refer to \cite{Franke2012}.} to the energy landscape in Derrida's Random Energy Model (REM) of spin 
glasses \cite{Derrida1980,Derrida1981}. For completeness we note that also the 
REM in an external magnetic field has an evolutionary analogue in the Rough Mt. Fuji (RMF) model
\cite{Franke2011,Franke2010}.

The mean number of local maxima of the HoC landscape can be obtained from a simple argument. 
A given genotype is a local maximum if its fitness value exceeds that of its $L$ neighbors,
which is true with probability $\frac{1}{L+1}$ by symmetry. Since there is a total
of $2^L$ genotypes, the expected value of the number $N_\mathrm{opt}$ of optima is \cite{Kauffman1987}
\begin{equation}
\label{HoC_opt}
\mathbb{E}(N_\mathrm{opt}) = \frac{2^L}{L+1}.
\end{equation}
The corresponding variance is \cite{Macken1989,Macken1991} 
\begin{equation}
\label{HoC_Var}
\mathrm{Var}(N_\mathrm{opt}) = \frac{2^L (L-1)}{2(L+1)^2},
\end{equation}
which implies that the coefficient of variation 
\begin{equation}
\label{CV_HoC}
C_V(N_\mathrm{opt}) = \sqrt{\frac{\mathrm{Var}(N_\mathrm{opt})}{\mathbb{E}(N_\mathrm{opt})^2}} = 
\frac{\sqrt{L-1}}{\sqrt{2}^{L+1}} 
\end{equation}
tends to zero for large $L$, i.e. the distribution of $N_\mathrm{opt}$ becomes increasingly localized
near its mean. In fact asymptotically the distribution is normal
\cite{Macken1989,Baldi1989}. For small $L$ the full distribution can
be obtained by exact enumeration, see Table \ref{Table1}.

\begin{table}
\caption{\label{Table1}Distribution of the number of fitness maxima in the HoC and constrained HoC
  models for $L=2$ and $L=3$. Note that the largest possible number of
maxima on the $L$-dimensional hypercube is $2^{L-1}$ \cite{Haldane1931}.}
\begin{tabular}{|c|c|c|c|}
\hline
&&&\\[-5pt] $L$	&	$N$	&	$\mathbb{P}(N_\mathrm{opt}^\mathrm{HoC} = N)$	&	$\mathbb{P}(N_\mathrm{opt}^\mathrm{cHoC} = N)$ \\ &&&\\[-5pt] \hline \hline
&&&\\[-5pt] 2	&	1	&	$\frac{2}{3}$	&	1	\\ &&&\\[-5pt] \hline
&&&\\[-5pt] 	&	2	&	$\frac{1}{3}$	&	0
\\ &&&\\[-5pt] \hline \hline
&&&\\[-5pt] 3	&	1	&	$\frac{3}{14} \approx 0.2143$	&	$\frac{1}{3} \approx 0.3333$	\\ &&&\\[-5pt] \hline
&&&\\[-5pt] 	&	2	&	$\frac{17}{28} \approx 0.6071$		&	$\frac{2}{5} = 0.4$	\\ &&&\\[-5pt] \hline
&&&\\[-5pt] 	&	3	&	$\frac{1}{7} = 0.1429$			&	$\frac{1}{5} = 0.2$	\\ &&&\\[-5pt] \hline
&&&\\[-5pt] 	&	4	&	$\frac{1}{28} \approx 0.0357$		&	$\frac{1}{15} \approx 0.0667$	\\ &&&\\ \hline
\end{tabular}
\end{table}

In a variant of the HoC model introduced in \cite{Kloezer2008,Carneiro2010} the global minimum
is constrained to be the antipodal sequence of the global maximum. This \textit{constrained} HoC
(cHoC) model can be implemented, e.g., 
by assigning fitness $F=1$ to $\sigma = (1,1,1,...,1)$, fitness $F = 0$ to 
$\sigma = (0,0,0,...,0)$ and random uniform fitness values in the interval $(0,1)$ to all other 
genotypes. Interestingly, the constraint does not change the expected number of fitness maxima, 
though it has a dramatic effect on the evolutionary accessibility of the landscape \cite{Franke2011,Hegarty2012},
see Sect.~\ref{Results_HoC} for further discussion. To see that eq.(\ref{HoC_opt}) is not
affected by the constraint, it is sufficient to note that the neighbors of the global minimum
have a slightly greater probability of being local maxima ($\frac{1}{L}$ instead of $\frac{1}{L+1}$),
which precisely compensates the reduction in the mean number of maxima which results from constraining
the antipode of the global maximum to be a minimum. This is true
provided the neighbors of the global maximum are not also neighbors of the
global minimum, i.e. for $L > 2$.

\subsection{Block model}

In the block model introduced by Perelson and Macken \cite{Perelson1995} the $L$ loci are grouped into $b$ disjoint sets (blocks) $B_1,\ldots,B_b$. Each block contributes an independent additive amount to the overall fitness of the genotype,
\begin{align}
\label{Fblock}
    F(\sigma) = \sum_{i=1}^b f_i(P_i \sigma)
\end{align}
where $P_i$ is the projector onto the subspace of $\mathbb H_2^L$ spanned by the loci in $B_i$. 
The value of $f_i$ depends only on the state of the loci in $B_i$. In the original version of the model the $f_i$ are drawn independently for each of the $2^{\vert B_i \vert}$ configurations of the 
loci in $B_i$, as in the HoC model, and we will adhere to this simple case in the following.
Similar to the HoC model, all properties of the model are then
manifestly independent of the distribution used to generate the random fitness values.  
However in principle the model can be extended to allow for any type
of fitness landscape within the blocks. In order to keep formulas simple we
will also assume all blocks to have the same size $m=\frac{L}{b}$. 
Most results may easily be generalized to varying block sizes.

To determine the mean number of local optima for the block model, we note that a genotype
is a local maximum of the fitness function (\ref{Fblock}) iff all projected configurations
$P_i \sigma$ are local maxima of the corresponding $f_i$. It follows that 
\begin{equation}
\label{Nopt_block}
N_\mathrm{opt} = \prod_{i=1}^b N_\mathrm{opt}^{(i)}
\end{equation}
where $N_\mathrm{opt}^{(i)} \geq 1$ is the number of maxima in block $i$. Since blocks are
independent, using (\ref{HoC_opt}) we obtain the expected number of maxima of the whole landscape as
\cite{Perelson1995} 
\begin{equation}
\label{block_opt}
\mathbb{E}(N_\mathrm{opt}) = [\mathbb{E}(N_\mathrm{opt}^{(i)})]^b = \frac{2^L}{(m+1)^b}.
\end{equation}
Similarly arbitrary moments of $N_\mathrm{opt}$ can be computed, and in particular the 
variance is given by \cite{Perelson1995}
\begin{equation}
\label{block_opt_Var}
\mathrm{Var}(N_\mathrm{opt}) = [\mathbb{E}(N_\mathrm{opt})]^2 
\left[ \left(1 + \frac{m-1}{2^{m+1}}\right)^b - 1 \right].
\end{equation}
While the expected number of optima (\ref{block_opt}) increases monotonically when the block size
$m$ is increased at fixed $L$, from $N_\mathrm{opt} = 1$ at $m=1$ to (\ref{HoC_opt}) for $m=L$, the 
variance (\ref{block_opt_Var}) is maximal at an intermediate value of $m$, and the coefficient
of variation is maximal at $m=2$ and $m=3$. At fixed $m$, $C_V$ increases exponentially with $L$, 
which implies that the 
distribution of the number of optima is very broad, in qualitative difference to the 
behavior of the HoC model.

\subsection{NK model}

The NK-model was introduced by Kauffman and coworkers \cite{Kauffman1989,Kauffman1993} to describe 
fitness landscapes with tunable ruggedness. In this model each locus in the genome contributes an additive amount to the total fitness of a given sequence. However the contribution of the $i$-th locus given by the real-valued function $f_i$  depends not only on the state of locus $i$ itself, but also on $k$ other loci $l_{i,1},\ldots,l_{i,k}$, called the neighbors of locus $i$. This implements epistatic interactions and enables one to model varying degrees of ruggedness depending on the parameter $k$. The total fitness
is then of the form
\begin{equation}
\label{NK}
F(\sigma) = \sum\limits_{i = 1}^L f_i(\sigma_i;\sigma_{l_{i,1}},\ldots,\sigma_{l_{i,k}}),
\end{equation}
where the values of the fitness contributions $f_i$ are taken to be identically distributed random variables drawn independently for each of the $2^{k+1}$ arguments. Common choices for the underlying
probability distribution are the uniform distribution or the standard normal distribution, and here 
we will always use the latter. The NK fitness landscape (\ref{NK}) includes the fully additive
landscape and the HoC model as limiting cases corresponding to $k=0$ and $k=L-1$, respectively.
From the perspective of spin glass physics, the NK-model can be viewed
as a superposition of diluted 
$p$-spin models \cite{Derrida1980,Derrida1981} with $p \leq k+1$ \cite{Stadler1999,Neidhart2013}.

Different genetic architectures can be implemented depending on how the neighbors of a locus are determined. There are various choices one might think of \cite{Weinberger1991,Fontana1993,Altenberg1997}. The most studied case is that of random neighbors (RN) in which the neighbors of each locus are drawn randomly with equal probability from the other loci. Another possible choice is the adjacent neighbors (AN) model in which the 
neighborhoods consist of $k+1$ consecutive loci along the sequence. To be specific, here we will take the 
neighbors of a locus in the AN model to be the $\lceil \frac{k}{2} \rceil$ loci preceding it and the 
$\lfloor \frac{k}{2} \rfloor$ loci succeeding it. 
In order to make this work the sequence is arranged in a circle.
It should be noted that the RN model contains two distinct sources of randomness, arising from the
choice of neighborhoods and the assignement of fitness values, respectively, whereas only the 
latter is present in the AN model. 

\begin{figure}
\centerline{\includegraphics[width=0.8\textwidth]{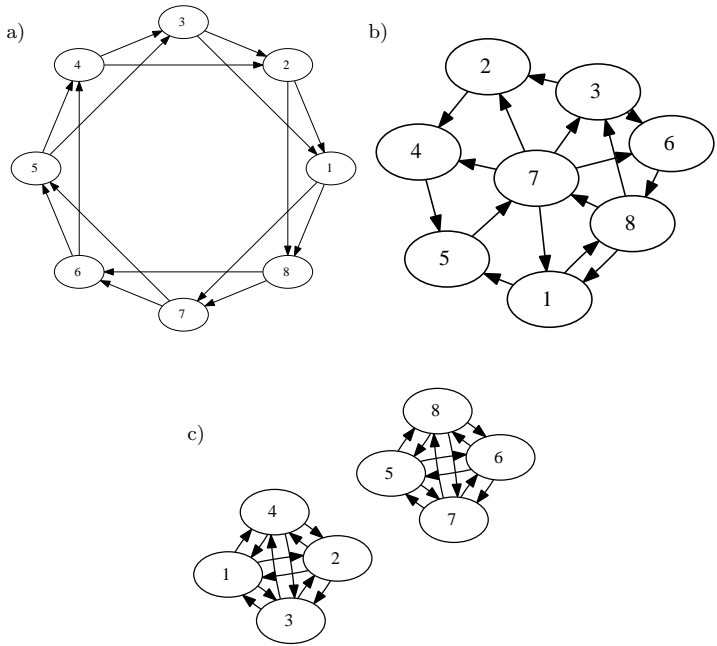}}
\caption{\label{Fig:NK} Example neighborhood graphs for the NK-model with $L=8$ loci. a) Adjacent neighborhoods (AN) with $k=2$. b) Random neighborhoods (RN) with $k=2$. c)  Block neighborhoods 
(BN) with $k=3$.}
\end{figure}

The different neighborhood choices  
can be represented as directed graphs over the set of loci, such that an edge directed from locus $i$ to locus $j$ exists if and only if the fitness contribution of locus $j$ depends on the state of locus $i$ (Fig.~\ref{Fig:NK}). Self-loops are not allowed since the dependence of $f_i$ on $\sigma_i$ is mandatory (but see \cite{Fontana1993,Neidhart2013,Altenberg1997} for versions of the model where this requirement is relaxed). The in-degree of each vertex is $k$, but the out-degree of vertices may vary,
e.g., as in the RN model. 
However the average out-degree must also be $k$ since all outgoing edges need to point to a vertex.
Within this framework
the block model (BN) of Perelson and Macken \cite{Perelson1995} is a special case of the NK model where the neighbors are chosen such that the neighborhood graph consists of $b$ components 
which are complete graphs and $k+1=m=L/b$ [Fig.~\ref{Fig:NK} c)].

\begin{figure}
\centerline{\includegraphics[width=\textwidth]{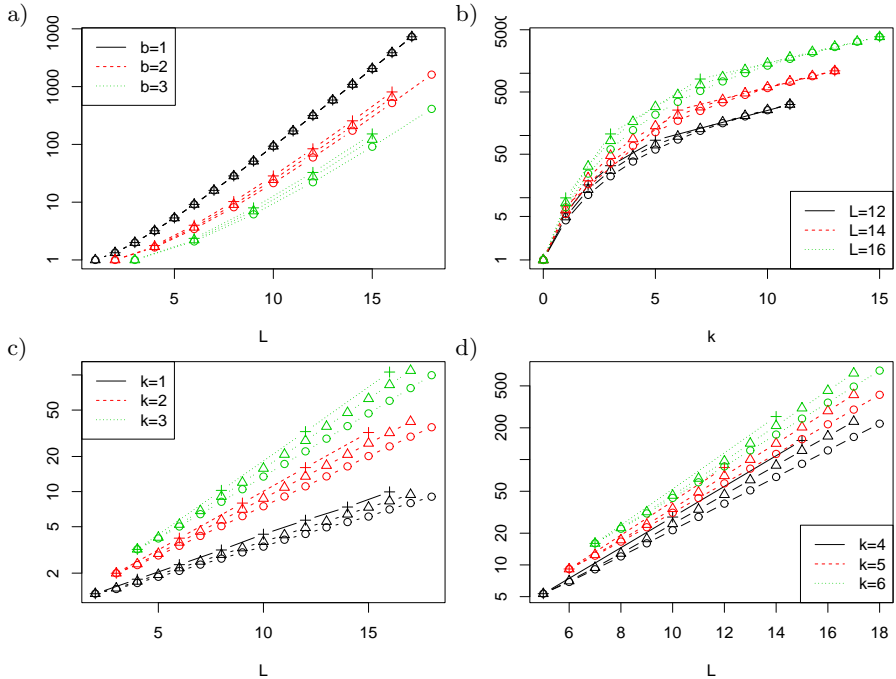}}
\caption{\label{Fig:CompOptMean} Mean number of local fitness maxima for the NK model with random neighborhood (RN, circles), adjacent neighborhood (AN, triangles) and block neighborhood (BN, crosses). a) Number of maxima as a function of $L$ for 
different values of $b=L/(k+1)$.  For $b=1$ the model
reduces to the HoC landscape and all versions are
equivalent. b) Number of maxima as a function of $k$ for different values
of $L$. c) Number of maxima as a function of $L$ for $k=1,2,3$. d) 
Number of maxima as a function of $L$ for $k=4,5,6$. Results for the block model (BN) are exact,
and simulation data for RN and AN neighborhoods were obtained from
$10^5$ ($10^4$) realizations per data point
for $L \leq 10$ ($L \geq 11$).}
\end{figure}

To what extent the choice of genetic architecture affects the properties of NK fitness landscapes 
is poorly understood. The two-point fitness correlation function is manifestly independent
of this choice \cite{Campos2002}, a statement that can be extended to the amplitude spectra 
obtained by Fourier transforming the landscape \cite{Neidhart2013}, but for other properties
such as the statistics of maxima the dependence on the structure of the neighborhoods is unknown. 
In this context it is instructive to compare the exact results for the block model reviewed in 
the previous subsection to available estimates for the number of fitness maxima in the NK model.
For fixed $k>0$ it has been established that the mean number of maxima grows exponentially 
with $L$, in the sense that \cite{Evans2002,Durrett2003}
\begin{equation}
\label{Max_asymp}
\lim_{L \to \infty} \frac{1}{L} \ln \mathbb{E}(N_\textrm{opt}) = \ln(2 \lambda_k)
\end{equation}
with a $k$-dependent constant $ \frac{1}{2} <  \lambda_k \leq 1$ that is expected
to also depend on the choice of neighborhoods and the underlying distribution from which the
fitness values are drawn. Comparing to eq.~(\ref{block_opt}) we see that the block model
expression for the $\lambda_k$ reads $\lambda_k^\textrm{block} = (k+2)^{-\frac{1}{k+1}}$.
Explicit results for the AN model with $k=1$ and various 
fitness distributions fall into the range $0.55463... \leq \lambda_1 \leq 0.5769536...$
\cite{Evans2002,Durrett2003}, which is remarkably close to (but slightly below) the block
model value $3^{-\frac{1}{2}} \approx 0.57735..$. Similarly the value $\lambda_2 = 0.611409...$
reported in \cite{Evans2002} for the AN model with an exponential fitness distribution is only
a few percent smaller than the block model value $4^{-\frac{1}{3}} \approx 0.62996...$.
This suggests that $\lambda_k^\textrm{block}$ may be an upper bound 
to $\lambda_k$ for any choice of neighborhoods.
A second class of rigorous results concerns the asymptotics when both $k$ and $L$ become large.
Under fairly general conditions it can be proved that for $L, k \to \infty$ \cite{Limic2004} 
\begin{equation}
\label{bothNK}
\ln \mathbb{E}(N_\textrm{opt}) - L \ln 2 \approx - \frac{L \ln k}{k},
\end{equation}
which also follows from the block model result (\ref{block_opt}).

\begin{figure}
\includegraphics[angle=-90,width=\textwidth]{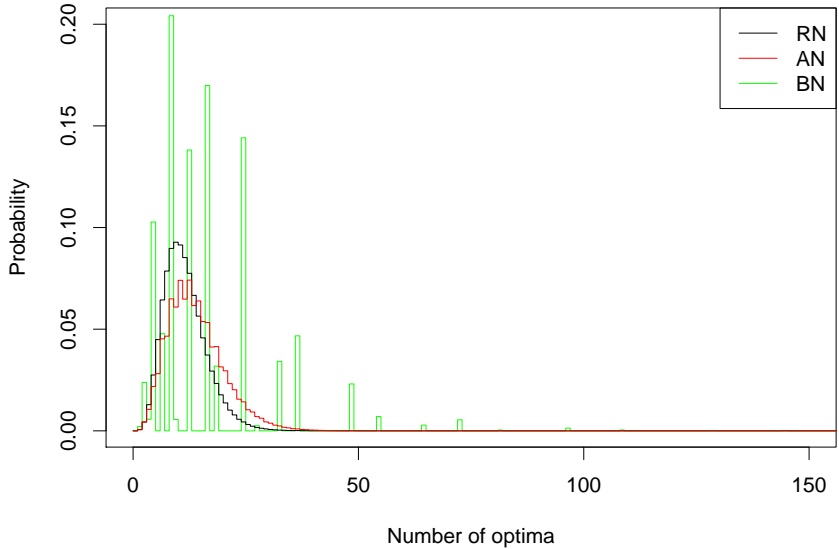}
\caption{\label{Fig:maxdist} Simulated distributions of the number of maxima for three different
versions of the NK-model with $L=12$ and $k=2$. Data were obtained from $10^6$ landscape realizations.}
\end{figure}

\begin{figure}
\centerline{\includegraphics[width=\textwidth]{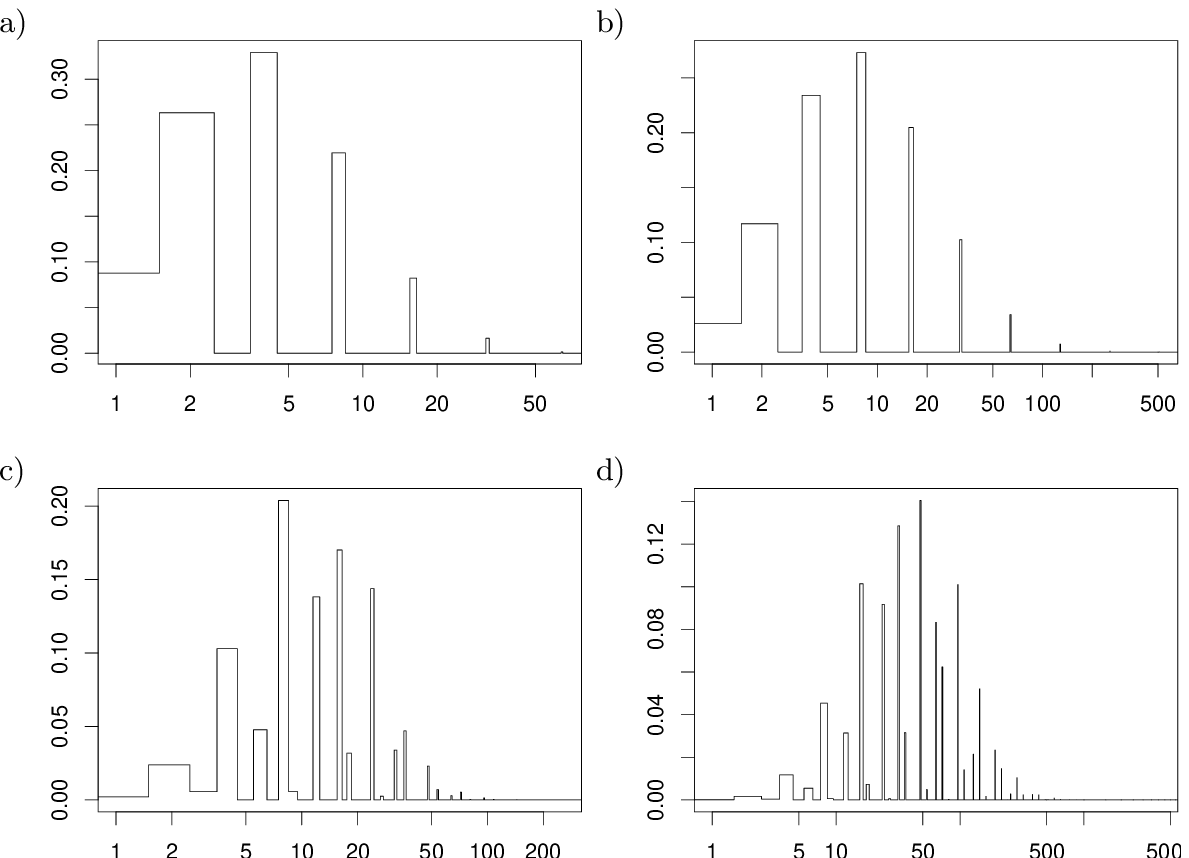}}
\caption{\label{Fig:BlockOptimaFull} Exact distribution of the number
  of maxima in the block model for a) $L=12$ and $m=2$, b) $L=18$ and $m=2$, c) $L=12$ and $m=3$ and d) $L=18$ and $m=3$. The number of maxima is shown in logarithmic scales in order to illustrate
the roughly log-normal shape of the distributions.}
\end{figure}

Taken together these 
observations indicate that the expected number of maxima in the NK-model depends only
weakly on the imposed genetic architecture, such that the block model provides a good
approximation to this quantity also for other versions of the
NK-model. This is illustrated in Fig.~\ref{Fig:CompOptMean}, which
compares the exact block model result (\ref{block_opt}) to numerical
data for the RN and AN models. 
Nevertheless, because of the specific multiplicative
structure of eq.~(\ref{Nopt_block}) the distribution of $N_\textrm{opt}$ in the block model
differs qualitatively from that in generic versions of the NK-model. 
As shown in Fig.~\ref{Fig:maxdist}, the RN- and AN-distributions
have a rather smooth appearance already for $L=12$, whereas the
corresponding BN-distribution features a pattern of discrete peaks,
see also Fig.~\ref{Fig:BlockOptimaFull}. In Figs.~\ref{Fig:maxdist}
and \ref{Fig:BlockOptimaFull} the block
sizes are $m=k+1=2$ or 3, and therefore the exact BN-distributions can be
generated directly from eq.~(\ref{Nopt_block}) using the corresponding distributions
for the HoC landscapes with $L=2$ and 3 given in Table \ref{Table1}.   
For larger values of $L$ and $m$ the envelopes of the distributions in Fig.~\ref{Fig:BlockOptimaFull} 
are seen to approach a log-normal shape, as might be expected from the multiplicative form of 
(\ref{Nopt_block}).

\begin{figure}
\centerline{\includegraphics[width=\textwidth]{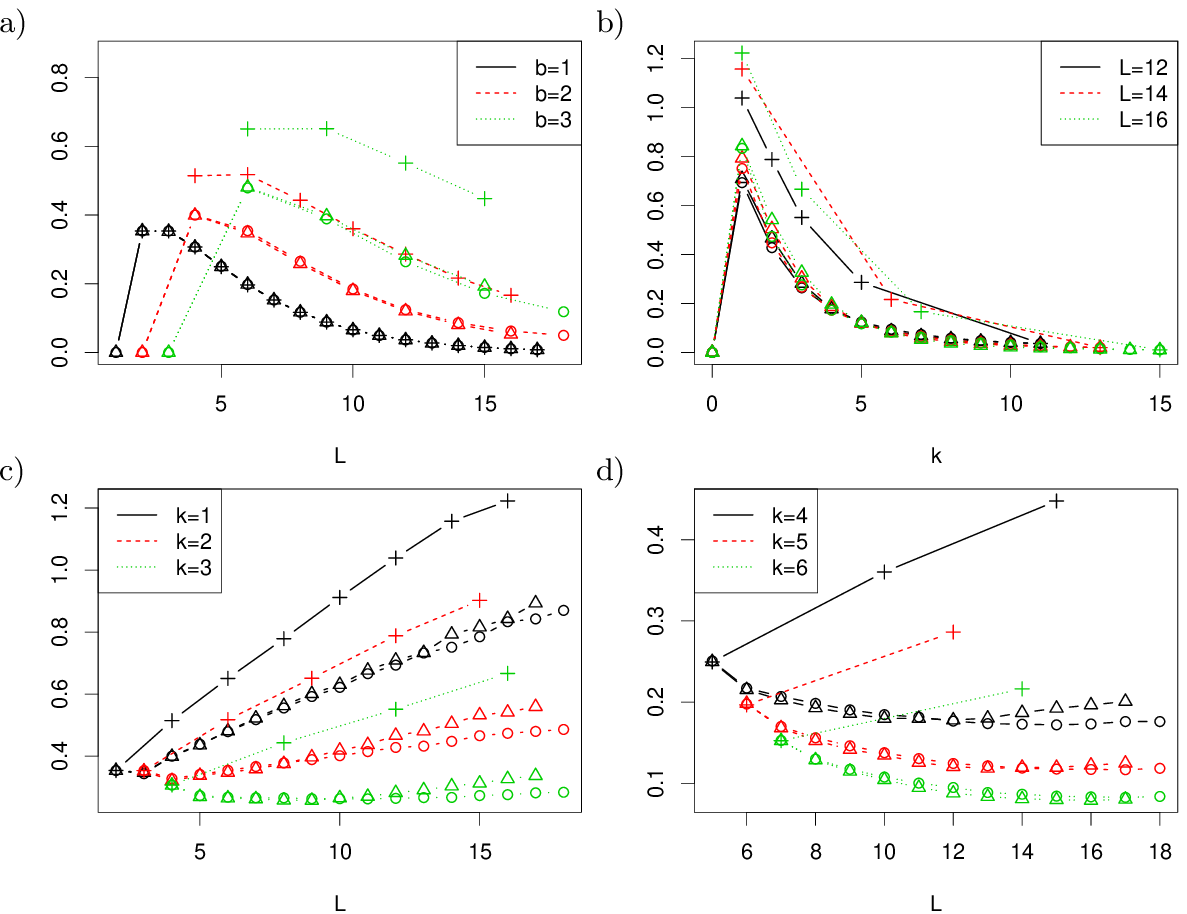}}
\caption{\label{Fig:CompOptCV} Coefficient of variation of the number
  of maxima for the NK model with random neighborhood (RN, circles), adjacent neighborhood (AN, triangles) and block neighborhood (BN, crosses) shown a) as a function of $L$ for  
different values of $b=L/(k+1)$, b) as a function of $k$ for different values
of $L$, c) as a function of $L$ for $k=1,2,3$ and d) 
as a function of $L$ for $k=4,5,6$. Results were obtained from simulations of $10^5$ ($10^4$) 
landscape realizations per data point for $L \leq 10$ ($L \geq 11$).}
\end{figure}

Figure \ref{Fig:CompOptCV} compares the coefficient of variation of
the number of maxima in the block model to the RN and AN versions of
the NK-model. In particular the data for $k \geq 4$ show a marked
qualitative difference between the models, in that $C_V$ grows with
sequence length for the block model while it appears to decrease for the other
versions (but note that $C_V$ may increase again at larger values of $L$). 
Thus, while the mean value of $N_\mathrm{opt}$ is rather
insensitive to the choice of neighborhoods, the fluctuations
in this quantity strongly reflect the genetic architecture of the
model. We will see below that similar statements can be made about the
distribution of selectively accessible pathways.

\section{Paths to the global maximum}
\label{Results}

\begin{figure}
\centering
\includegraphics[angle=-90,width=\textwidth]{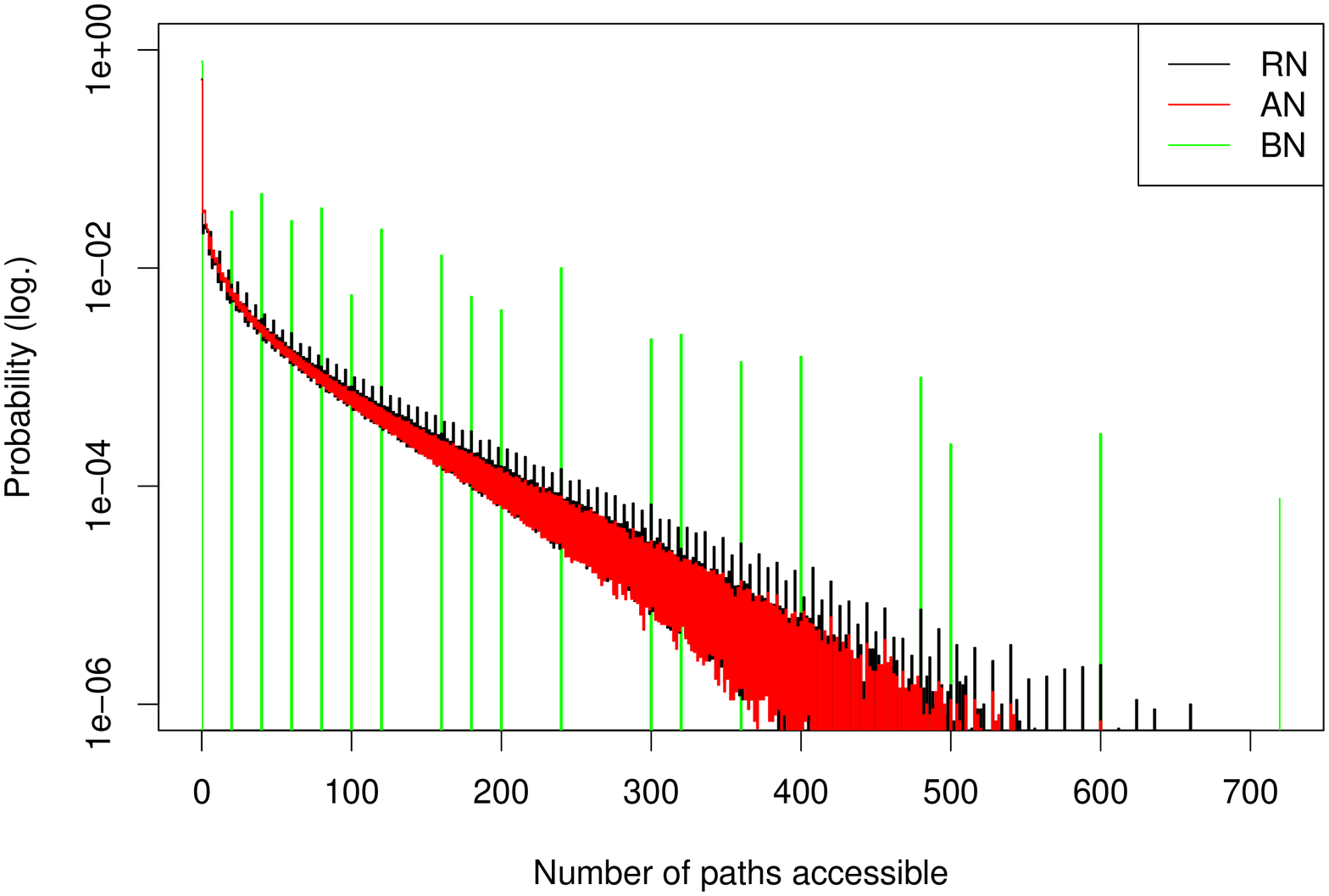}
\caption{\label{Fig:PathsExample}Simulated distributions of the number of accessible paths for different 
versions of the NK-model with $L=6$ and $k=2$. Data were obtained from $10^7$ landscape realizations. This is a line histogram similar to Fig.~\ref{Fig:BlockOptimaFull}. 
The apparent width of the black and red lines
reflects the high frequency variations in the probability between neighboring values. 
These are not due to limitations of the simulation but are a feature
of the model. While the probability for all path counts up to at least 400 is bounded away from
zero for RN and AN landscapes, the set of possible path counts in the BN model is very sparse,
that is, the probability is exactly zero in the gaps visible in the BN distribution.
For a discussion of similar gaps in the RN distribution we refer to \cite{Franke2012}.}
\end{figure}

In the SSWM regime the population generates and possibly fixes mutations one by one,
and transitions involving several mutations at a time are not possible.
Nonetheless there are many possible paths through the hypercube that connect pairs of genotypes.
In the following we will only consider paths of minimal length. In
this case any permutation of the mutations necessary to transform one
genotype into the other is a valid pathway, resulting in $d!$ possible paths
connecting genotypes at Hamming distance $d$. Following earlier work
\cite{Weinreich2005,Franke2011,Carneiro2010,Franke2012,Hegarty2012,Nowak2013,Berestycki2013,Roberts2013} 
we will focus specifically on paths that end at the global
fitness maximum of the landscape $\Omega \in \mathbb H_2^L$ and start
at the antipodal node $\bar{\Omega} = \Delta_{\{1\ldots L\}} \Omega$.
Each path $p$ is then uniquely defined as one of the $L!$ permutations of
all loci, where the order of loci corresponds to the order in which
mutations occur, $p = (p_1,\ldots,p_L) \in \mathrm{Perm}_L$ \cite{Gokhale2009}.
Under strong selection each introduced mutation has to increase
fitness in order to prevail in the population. A path through the
fitness landscape is therefore called \textit{selectively accessible} if and only if
each step increases fitness, that is, iff
$F(\Delta_{\{p_1,\ldots,p_i\}}\bar{\Omega}) > F(\Delta_{\{p_1,\ldots,p_{i-1}\}}\bar{\Omega})$ for all $i$
\cite{Weinreich2005}.

The object of interest in this section is the distribution
of the number of selectively accessible paths
$N_\mathrm{p}$ to the global maximum, a random variable taking values
between 0 and $L!$. In Fig. ~\ref{Fig:PathsExample} we show path
distributions for the block model and two versions of the NK-model
(see \cite{Franke2011,Franke2012} for further numerical examples). While the distributions for
the AN and RN models look reasonably continuous, in the block model 
only a discrete set of path numbers is allowed. 
As we will see below in Sect.~\ref{Results_BN}, the allowed numbers are in fact 
integer multiples of a constant arising from the block structure.

Of particular importance for the characterization of the statistics of
accessible paths is the probability
$\mathbb{P}(N_\mathrm{p} > 0)$ of finding at least one such path, a
quantity that has been introduced in earlier work as an overall
measure of landscape ruggedness \cite{Franke2011,Carneiro2010} and
that will be referred to as the
\textit{accessibility} of the fitness 
landscape in the following. Since the paths under consideration
are those that span the entire hypercube, asking for the
probability of their existence is obviously akin to a percolation
problem \cite{Nowak2013}. 

Intuitively one expects that the accessibility should be related 
to the average number of accessible paths $\mathbb{E}(N_\mathrm{p})$, which is usually
easier to compute than $\mathbb{P}(N_\mathrm{p} > 0)$. 
Specifically, it was conjectured in \cite{Franke2011} that 
$\mathbb{P}(N_\mathrm{p} > 0) \to 1$ for $L \to \infty$ whenever $\mathbb{E}(N_\mathrm{p})$
grows without bound in this limit. In order to sharpen this intuition it is instructive 
to examine the inequalities
\begin{equation}
\label{2ndmoment}
\frac{\mathbb{E}(N_\mathrm{p})^2}{\mathbb{E}(N_\mathrm{p}^2)} \leq 
\mathbb{P}(N_\mathrm{p} > 0) \leq \mathbb{E}(N_\mathrm{p})
\end{equation}
which hold for any non-negative, discrete random variable \cite{Alon2000} and have been employed in several recent studies of evolutionary accessibility \cite{Hegarty2012,Nowak2013,Roberts2013}.
Two general conclusions can be drawn from (\ref{2ndmoment}). First, if $\lim_{L \to \infty}
\mathbb{E}(N_\mathrm{p}) = 0$, then the same holds true for $\mathbb{P}(N_\mathrm{p} > 0)$. 
Second, if $\mathbb{E}(N_\mathrm{p})$ diverges for $L \to \infty$ \textit{and} if the distribution 
of path numbers is sufficiently centered around the mean such that  
$\lim_{L \to \infty} \mathbb{E}(N_\mathrm{p})^2/\mathbb{E}(N_\mathrm{p}^2) = 1$, then indeed
$\mathbb{P}(N_\mathrm{p} > 0) \to 1$ in the limit. The latter scenario has been established
in \cite{Hegarty2012} for the constrained HoC and the RMF models. Below we will see that the
block model displays a different and somewhat counterintuitive behavior, in that
$\mathbb{E}(N_\mathrm{p})$ increases rapidly with $L$ but nevertheless 
$\lim_{L \to \infty} \mathbb{P}(N_\mathrm{p} > 0) = 0$. As can be read off from the first
inequality in (\ref{2ndmoment}), this is only possible if $N_\mathrm{p}$ remains a strongly
fluctuating quantity, such that $\mathbb{E}(N_\mathrm{p}^2) \gg \mathbb{E}(N_\mathrm{p})^2$
for large $L$ (see also Sect.~\ref{Pathfluctuations}).

\subsection{HoC model}
\label{Results_HoC}

A path to the global maximum in the HoC model consists of $L+1$
independent and identically distributed fitness values, the last one
of which is known to be larger than all the others. The probability
for the remaining $L$ values to be in ascending order is then
$\frac{1}{L!}$ by symmetry, and the expected number of paths 
is \cite{Franke2011}
\begin{equation}
\label{HoC_Paths}
\mathbb{E}(N_\mathrm{p}^\mathrm{HoC}) = 1
\end{equation}
independent of $L$, which does not yield a nontrivial upper bound on the accessibility
through (\ref{2ndmoment}). Hegarty and Martinsson
\cite{Hegarty2012} have proved that the accessibility in fact tends to
zero asymptotically as 
\begin{equation}
\label{Access_HoC}
\mathbb{P}(N_\mathrm{p}^\mathrm{HoC} > 0) \sim \frac{\ln{L}}{L}.
\end{equation}
Together these results imply that, conditioned on accessible
realizations with $N_\mathrm{p} > 0$, the expected number of paths
grows with $L$ as 
\begin{equation}
\label{HoC_Paths_IfAccess}
\mathbb{E}(N_\mathrm{p}^\mathrm{HoC} | N_\mathrm{p}^\mathrm{HoC} > 0) \sim \frac{L}{\ln{L}},
\end{equation}
but even in that case only a vanishing fraction of all $L!$ paths will
be accessible.

Turning to the constrained HoC model where $\bar{\Omega}$ is
constrained to be the global fitness minimum, the combinatorial
argument leading to (\ref{HoC_Paths}) shows that \cite{Franke2011} 
\begin{equation}
\label{cHoC_Paths}
\mathbb{E}(N_\mathrm{p}^\mathrm{cHoC}) = L
\end{equation}
and accessibility increases dramatically, in the sense that \cite{Hegarty2012}
\begin{equation}
\label{Access_cHoC}
\lim_{L \to \infty} \mathbb{P}(N_\mathrm{p}^\mathrm{cHoC} > 0) = 1,
\end{equation}
see \cite{Franke2011} for numerical evidence pointing in this
direction. Moreover, it is shown in \cite{Hegarty2012} that the
variance of the number of paths in the cHoC model behaves
asymptotically as $\mathrm{Var}(N_\mathrm{p}^\mathrm{cHoC}) \approx 4
L^2$, and correspondingly the coefficient of variation saturates at a
value of 2. The results of \cite{Hegarty2012} for the cHoC model can be adapted
to show that for the unconstrained model  
\begin{equation}
\label{HoC_Paths_Var}
\mathrm{Var}(N_\mathrm{p}^\mathrm{HoC}) \approx 2L, 
\end{equation}
which implies that the cofficient of variation grows with $L$ as
$\sqrt{2L}$.
For completeness we note that the exact value of the variance is
$\mathrm{Var}(N_\mathrm{p}^\mathrm{HoC}) = \frac{2}{3}$ for $L=2$ and
$\mathrm{Var}(N_\mathrm{p}^\mathrm{HoC}) = \frac{19}{10}$ for $L=3$,
as can be derived from the full distribution displayed in Table
\ref{Table2}.

\begin{table}
\caption{\label{Table2} Exact distribution of the number of accessible paths in the HoC and constrained HoC
  models for $L=2$ and $L=3$.}
\begin{tabular}{|c|c|c|c|}
\hline
&&&\\[-5pt] $L$	&	$N$	&	$\mathbb{P}(N_\mathrm{p}^\mathrm{HoC} = N)$	&	$\mathbb{P}(N_\mathrm{p}^\mathrm{cHoC} = N)$ \\ &&&\\[-5pt] \hline \hline
&&&\\[-5pt] 2	&	0	&	$\frac{1}{3}$	&	0	\\ &&&\\[-5pt] \hline
&&&\\[-5pt] 	&	1	&	$\frac{1}{3}$	&	0	\\ &&&\\[-5pt] \hline
&&&\\[-5pt] 	&	2	&	$\frac{1}{3}$	&	1	\\ &&&\\[-5pt] \hline \hline

&&&\\[-5pt] 3	&	0	&	$\frac{113}{210} \approx 0.5381$	&	$\frac{1}{15} \approx 0.0666...$	\\ &&&\\[-5pt] \hline
&&&\\[-5pt] 	&	1	&	$\frac{51}{280} \approx 0.1821$		&	$\frac{13}{120} \approx 0.108333...$	\\ &&&\\[-5pt] \hline
&&&\\[-5pt] 	&	2	&	$\frac{11}{84} \approx 0.1310$		&	$\frac{13}{60} \approx 0.21666...$	\\ &&&\\[-5pt] \hline
&&&\\[-5pt] 	&	3	&	$\frac{31}{420} \approx 0.0738$		&	$\frac{13}{60} \approx 0.21666...$	\\ &&&\\[-5pt] \hline
&&&\\[-5pt] 	&	4	&	$\frac{1}{20} = 0.05$			&	$\frac{13}{60} \approx 0.21666...$	\\ &&&\\[-5pt] \hline
&&&\\[-5pt] 	&	5	&	$\frac{13}{840} \approx 0.0155$		&	$\frac{13}{120} \approx 0.108333...$	\\ &&&\\[-5pt] \hline
&&&\\[-5pt] 	&	6	&	$\frac{1}{105} \approx 0.0095$
&	$\frac{1}{15} \approx 0.0666...$	\\ &&&\\  \hline
\end{tabular}
\end{table}

\subsection{Block model}
\label{Results_BN}

Consider a block landscape with $b$ blocks $B_1,\ldots,B_b$ of size $m = \frac{L}{b}$. A mutation $\Delta_{i_1}$ mutating a locus ${i_1}$ in block $B_{j_1}$ will only change the fitness contribution of this block $f_{j_1}$,
\begin{equation}
  \Delta_{i_1} F(\sigma) = \Delta_{i_1} f_{j_1}(P_{j_1}\sigma).
\end{equation}
A subsequent mutation $i_2$ in a different block $B_{j_2}$ generates the fitness change
\begin{equation}
  \Delta_{i_2} F(\Delta_{i_1} \sigma) = \Delta_{i_2} f_{j_2}(P_{j_2}\Delta_{i_1}\sigma)
\end{equation}
which, since $i_1$ does belong to block $B_{j_2}$, simplifies to
\begin{equation}
    \Delta_{i_2} F(\Delta_{i_1}\sigma) = \Delta_{i_2} f_{j_2}(P_{j_2}\sigma)
\end{equation}
Hence the order in which two loci are mutated is irrelevant to the accessibility if the two loci are not part of the same block and are mutated directly one after another.
Introducing the indicator function
\begin{equation}
X(p) = \left\{\begin{aligned}
              1	&\;\;&	\text{if path \textit{p} is accessible} \\
	      0 &\;\;&	\text{if path \textit{p} is not accessible}
             \end{aligned}\right.
\end{equation}
this property reads $X((\ldots,i_1,i_2,\ldots)) = X((\ldots,i_2,i_1,\ldots))$.

Consider now a path $p = (p_1,\ldots,p_L)$. Switching two adjacent elements of the path will not change the accessibility if they do not share a block.
It is therefore possible to reorder the path in the form $\bar p = (\bar p_1, \ldots, \bar p_L)$ such that $\{\bar p_{(i-1)m+1}, \ldots, \bar p_{im}\} = B_i$ for all $i$ and $X(p) = X(\bar p)$.
For each such ordered path there are $\frac{L!}{m!^b}$ original paths reducing to it in the way described. The number of accessible paths on the block landscape therefore has to be an integer multiple of $\frac{L!}{m!^b}$.
Note that this feature of the block model does not depend on the
blocks consisting of HoC landscapes. The combinatorial factor is only
determined by the block structure and will be present in all
fitness landscapes composed of independent sets of loci.

The ordered path $\bar p$ can be divided into $b$ subpaths operating
on each block seperately. Steps in other blocks do not influence the
accessibility of the subpaths in a given block $B_i$. It is thus possible to
write the number of paths on the block landscape as the product of the number of paths in each block,
\begin{equation}
\label{BlockProduct}
 N_\mathrm{p}^\mathrm{BN} = \frac{L!}{m!^b}\prod\limits_{i=1}^b N_\mathrm{p}^{(i)},
\end{equation}
in close analogy to the corresponding relation (\ref{Nopt_block}) for the number
of maxima. 
The end point of a subpath $\Delta_{B_i} P_i \bar{\Omega}$ is also the
global maximum of the block landscape $f_i$, since $F$ is the sum of independent blocks.
Therefore the distribution of the number of paths to the global
maximum can be derived from the distribution of the number of paths to
the global maximum of the blocks according to
\begin{equation}
\label{Block_Paths_Full}
	\mathbb{P}(N_\mathrm{p}^\mathrm{BN} = N) = \left\{
		\begin{aligned} & 
			\sum\limits_{D_b(z)} \; \prod\limits_{i=1}^b
                        \mathbb{P}(N_\mathrm{p}^\mathrm{HoC(\textit{m})}
                        = n_i)  \;\;\; \textrm{if} \;\;\;  z = \frac{m!^b}{L!}\cdot N \in \mathbb{N}_0 \\
			& \;\;\; 0 \;\;\;	\textrm{else},
		\end{aligned}
		\right.
\end{equation}
where $D_b(z) = \{(n_1,\ldots,n_b) \in
\mathbb{N}_0^b\;|\;\prod_{i=1}^b n_i = z\}$ is the set of all ordered
decompositions of the non-negative integer $z$ into a product of $b$
non-negative integer factors and $N_\mathrm{p}^{\mathrm{HoC}(m)}$ is the
number of accessible paths in a HoC landscape of size $m$.
From this general relation together with the result
(\ref{HoC_Paths}) for the HoC model the following
expressions for the statistics of accessible paths in the block model emerge:
\begin{equation}
\label{Block_Paths}
\mathbb{E}(N_\mathrm{p}^\mathrm{BN}) = \frac{L!}{m!^b}
\end{equation}
\begin{equation}
\label{Block_Paths_Var}
\mathrm{Var}(N_\mathrm{p}^\mathrm{BN}) = \frac{L!^2}{m!^{2b}} \left(\mathbb{E}[(N_\mathrm{p}^{\mathrm{HoC}(m)})^2]^b - 1\right),
\end{equation}
\begin{equation}
\label{Block_Paths_CV}
C_V(N_\mathrm{p}^\mathrm{BN}) = \sqrt{\mathbb{E}[(N_\mathrm{p}^{\mathrm{HoC}(m)})^2]^b - 1},
\end{equation}
\begin{equation}
\label{Block_Paths_Access}
\mathbb{P}(N_\mathrm{p}^\mathrm{BN} > 0) = \left(\mathbb{P}(N_\mathrm{p}^{\mathrm{HoC}(m)} > 0)\right)^b.
\end{equation}
All of these results easily carry over to variations in which the
block landscapes are not of HoC type, however in the following we continue to assume HoC blocks.

\begin{figure}
    \centering
    \includegraphics[angle=-90,width=0.70\textwidth]{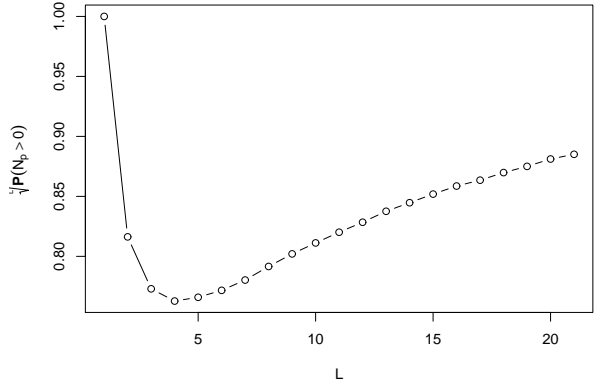}
    \caption{\label{Fig:AccessHoCRoot}Numerical estimates of the quantity $\mu_L = \sqrt[L]{\mathbb{P}(N_\mathrm{p}^{\mathrm{HoC}(L)} > 0)}$ 
as a function of $L$.}
\end{figure}

\subsubsection{Accessibility}

It follows from (\ref{Block_Paths_Access}) that the accessibility in
the block model always tends to zero, so block landscapes with high
$L$ almost surely do not have any path to the global maximum. 
In this regard there is no difference to the HoC model. However in the block model
accessibility tends to zero much faster. For fixed block size $m$ the decrease is
exponential in $L$, whereas for a fixed number of blocks $b$ the HoC
asymptotics  (\ref{Access_HoC}) implies that
$\mathbb{P}(N_\mathrm{p}^\mathrm{BN} > 0) \sim  (\ln L/L)^b$, which is smaller than (\ref{Access_HoC}) for any $b > 1$.
Since $b = L/m$, eq.~(\ref{Block_Paths_Access}) implies that accessibility at constant $L$ is governed by the quantity 
\begin{equation}
\label{root}
\mu_m \equiv \sqrt[m]{\mathbb{P}(N_\mathrm{p}^{\mathrm{HoC}(m)} > 0)}
\end{equation} 
defined such that $\mathbb{P}(N_\mathrm{p}^\mathrm{BN} > 0) = \mu_m^L$. 
By construction $\mu_1 = 1$, and according to the asymptotics (\ref{Access_HoC}) $\mu_m$ approaches unity from below
for large $m$ because $\lim_{m\rightarrow\infty} (\ln{m}/m)^\frac{1}{m} = 1$. 
It follows that $\mu_m$ is minimal at an intermediate block size, which turns out to be $m = 4$, see 
Fig.~\ref{Fig:AccessHoCRoot}. At $m=4$ the block model thus displays minimal accessibility.

\begin{figure}
\centerline{\includegraphics[width=\textwidth]{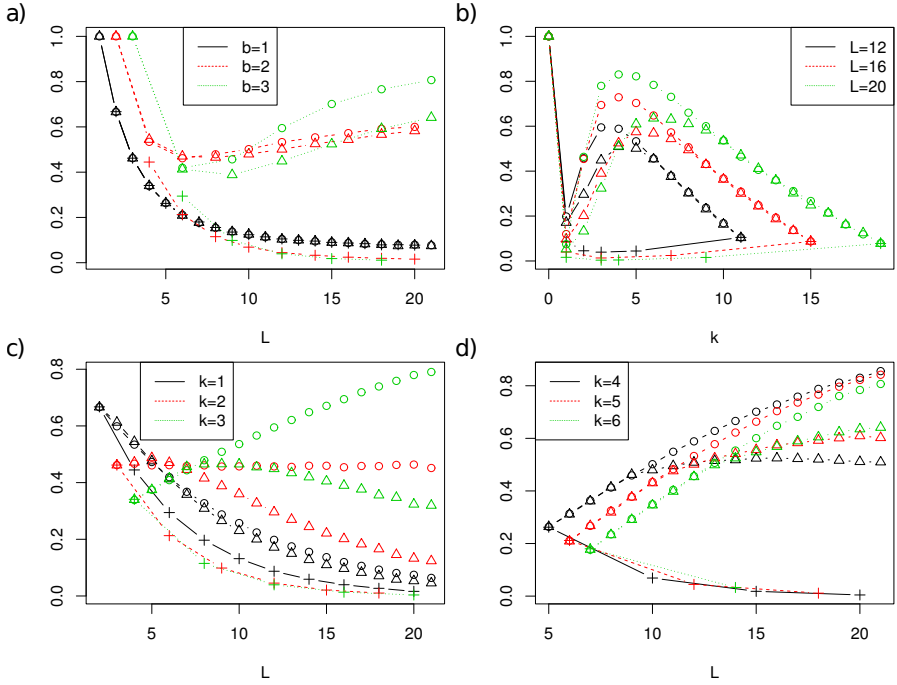}}
\caption{\label{Fig:CompAccess} Accessibility $\mathbb{P}(N_\mathrm{p} > 0)$ for the NK model with random neighborhood (RN, circles), adjacent neighborhood (AN, triangles) and block neighborhood (BN, crosses). a) Accessibility
as a function of $L$ at fixed block number $b$. For $b = 1$ all models are equivalent. For $b > 1$ the block model shows
monotonically decreasing accessibility which falls below the HoC value ($b = 1$) with increasing $L$, whereas for the RN and BN
models accessibility displays a minimum and increases for large $L$. b) Accessibility for fixed $L$ as a function of $k$. 
Block model data show a minimum at $k = m-1 = 3$, whereas RN and AN models display a maximum. For $k > L/2$ RN and AN
data are essentially indistinguishable. c) Accessibility as a function of $L$ for fixed $k = 1,2,3$. Block model data
decrease monotonically while the RN model displays a transition between decreasing accessibility for $k=1$ to 
increasing accessibility for $k=3$. d) Same as c) for
$k=4,5,6$. Results were obtained from simulations of $10^5$ landscape
realizations per data point.}
\end{figure}

Figure ~\ref{Fig:CompAccess} shows the comparison of evolutionary accessibility for the BN, AN and RN models. 
For constant $b$ [Fig.~\ref{Fig:CompAccess} a)] 
there is a significant difference between the behavior of HoC/BN models and AN/RN models. While the accessibility
in the HoC model and block model is monotonically falling, both the RN and AN model exhibit a minimum in the accessibility followed
by an increase for large $L$.
For constant $L$ the block model's minimal accessibility at $k = m-1 = 3$ is recognizable in 
Fig.~\ref{Fig:CompAccess} b). Interestingly, the AN and RN models display a reverted behavior with a maximum accessibility at intermediate
$k$. This figure also shows that the accessibility values for the RN and AN models are numerically indistinguishable for $k > L/2$
while important differences arise for smaller $k$, see also Figs.~\ref{Fig:CompAccess} c) and d). 
Compared to the HoC and block model the AN and RN models are surprisingly accessible even for high $L$. 
While it is virtually impossible to find a block landscape with accessible paths for
$L=16$, the AN and RN landscapes of that size have a chance of more than 50\% to be accessible for suitable values of $k$.

The comparison of different models at 
constant $k$ in Figs.~\ref{Fig:CompAccess} c), d) shows that the RN and AN models behave qualitatively similar to the block model 
for $k=1$, but differ strongly from the block model and from each other for $k \geq 2$.  
While the AN data generally seem to display a maximum followed by decreased accessibility for larger $L$, 
the accessibility in the RN model remains nearly independent of $L$ for $k=2$ and increases monotonically with $L$
for $k \geq 3$. The transition in accessibility at $k=2$ for the RN model was already observed and discussed in 
\cite{Franke2012}, but here we see that the behavior in the AN model appears to be qualitatively different.

\begin{figure}
\centerline{\includegraphics[width=\textwidth]{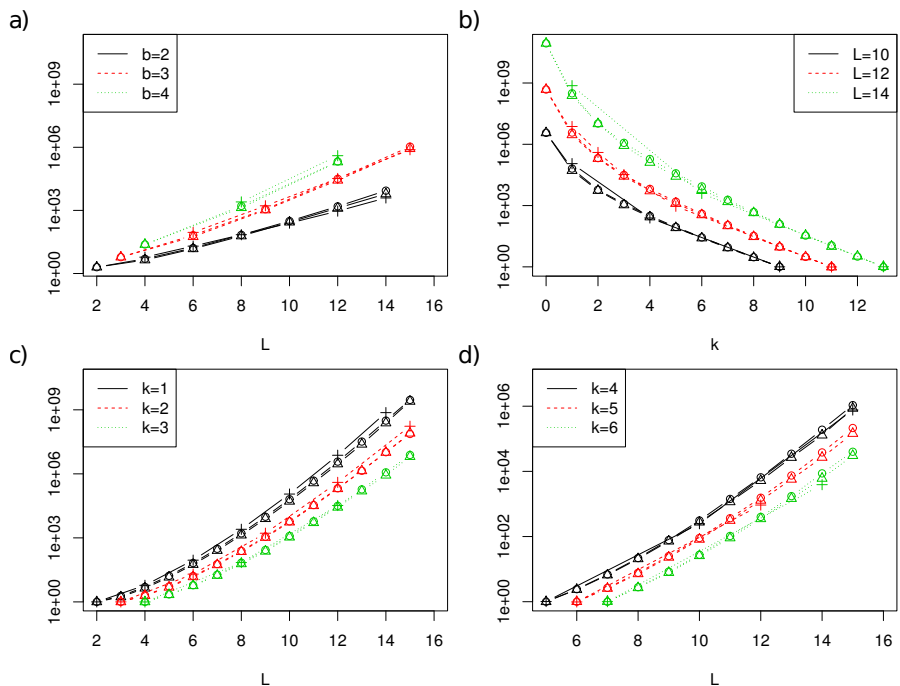}}
\caption{\label{Fig:CompPathsMean} Mean number of accessible paths for the NK model with random neighborhood (RN, circles), adjacent neighborhood (AN, triangles) and block neighborhood (BN, crosses) shown 
a) as a function of $L$ for different values of $b=L/(k+1)$, b) as a function of $k$ for different values
of $L$, c) as a function of $L$ for $k=1,2,3$ and d) 
as a function of $L$ for $k=4,5,6$. Results were obtained from simulations of $10^5$ ($10^4$) 
landscape realizations per data point for $L \leq 13$ ($L \geq 14$).}
\end{figure}

\subsubsection{Mean number of paths}

The mean number of paths (\ref{Block_Paths}) in the block model equals its first non-vanishing path count greater than zero 
which is a property inherited from the HoC model.
Asymptotically for large $L$ the mean behaves as
\begin{equation}
\label{Block_Paths_Asym}
\begin{split}
m = \text{const.}:\;\; &\mathbb{E}(N_\mathrm{p}^{\mathrm{BN}}) \approx \sqrt{2\pi}L^{L+\frac{1}{2}}\left(e\sqrt[m]{m!}\right)^{-L}, \\
b = \text{const.}:\;\; &\mathbb{E}(N_\mathrm{p}^{\mathrm{BN}}) \approx \left(\frac{1}{\sqrt{2\pi L}}\right)^{b-1} b^{L+\frac{b}{2}}.
\end{split}
\end{equation}
For constant block size $m$ the mean increases asymptotically faster than for constant block number $b$. 
Nonetheless, even for constant $b>1$ the mean path number on the block landscape increases
nearly exponentially and therefore much faster than the mean on HoC landscapes conditioned to be accessible, see 
eq.~(\ref{HoC_Paths_IfAccess}).

This behavior does not appear to be unique to the block model. In fact, simulation results shown in Fig.~\ref{Fig:CompPathsMean} 
suggest that the mean number of accessible paths in all versions of the NK-model is rather similar. The formula 
(\ref{Block_Paths}) derived above might therefore be useful for estimating the mean for these other variants of the NK model. 
A consistent ordering between the AN, RN and BN models is however not recognizable: 
While for small $k$ the mean for the block model is highest,
it becomes lowest in the regime of large $L$ and $k$.

\begin{figure}
\centerline{\includegraphics[width=\textwidth]{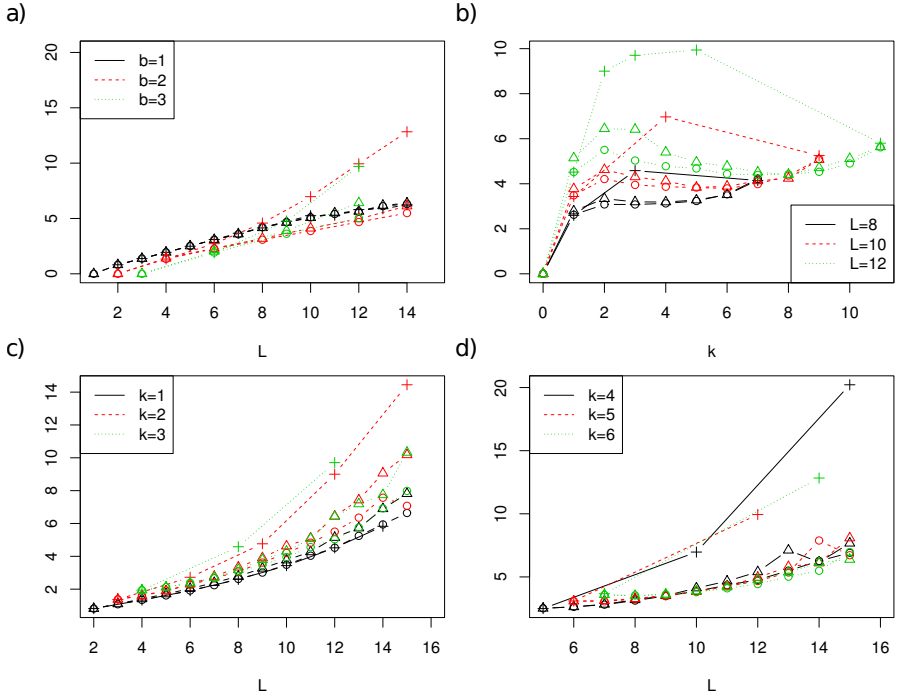}}
\caption{\label{Fig:CompPathsCV} Coefficient of variation of the number of
  paths for the NK model with random neighborhood (RN, circles), adjacent neighborhood (AN, triangles) and block neighborhood (BN, crosses). a) $C_V$ as a
  function of $L$ at fixed block size $b=L/(k+1)$; data are plotted on
  double-logarithmic scales to facilitate the comparison with the
  asymptotic prediction (\ref{Block_Paths_CV_Asym}). b) $C_V$ as a
function of $k$ at fixed $L$; note that all models coincide for $k=0$
(additive fitness landscape) and $k=L-1$ (HoC model). c) $C_V$ as a
function of $L$ at fixed $k=1,2,3$; d) same as c) for
$k=4,5,6$. Results were obtained from simulations of $10^5$ landscape
realizations per data point.} 
\end{figure}

\subsubsection{Fluctuations of the number of paths}
\label{Pathfluctuations}

To characterize the fluctuations in the number of accessible paths we consider the coefficient of variation 
$C_V(N_\mathrm{p})$. For the block model, the relation (\ref{Block_Paths_CV}) shows that $C_V$ increases 
exponentially with $L$ for fixed $m$, while for constant $b$ the asymptotic result (\ref{HoC_Paths_Var}) for the HoC model
implies that 
\begin{equation}
\label{Block_Paths_CV_Asym}
C_V(N_\mathrm{p}^\mathrm{BN}) \approx (2L)^\frac{b}{2}
\end{equation} 
for large $L$. 
Although the distribution of paths becomes increasingly broader with increasing $L$ also in the HoC model,
the increase of $C_V$ is thus seen to be much faster in the block model, especially for constant $m$.

The simulation results for $C_V$ displayed in Figure~\ref{Fig:CompPathsCV} a) show that the asymptotics
(\ref{Block_Paths_CV_Asym}) is attained only for sequence lengths
substantially larger than $L=10$, which are beyond the reach of our
simulations. The coefficient of variation for the block model is seen
to increase faster with $L$ for larger $b$, but even for $L=12$ the
ordering of the data points is not yet consistent with the asymptotic
behavior, in that $C_V$ is slightly larger for $b=2$ than for $b=3$.
 
The path number fluctuations in the RN and AN models are generally
smaller than in the BN model, with the exception of $k=1$,  where the
block model $C_V$ is very close to the value for the RN model, see
Fig.~\ref{Fig:CompPathsCV} b). This figure shows that the dependence
of $C_V$ on $k$ is generally non-monotonic, with a maximum attained at 
an intermediate value of $k$. The $L$-dependence of $C_V$ at fixed $k$
is shown in Figs.~\ref{Fig:CompPathsCV} c) and d). While all models
behave similarly for small $k=1,2,3$, at larger $k=4,5,6$ the increase
of $C_V$ is markedly steeper for the block model than for the other
models. At larger values of $k$ the RN and AN curves develop a
minimum in $L$ which is followed by a rapid increase (not shown).

\subsubsection{Exact distribution of the number of paths for small blocks}

\begin{figure}
\centerline{\includegraphics[width=\textwidth]{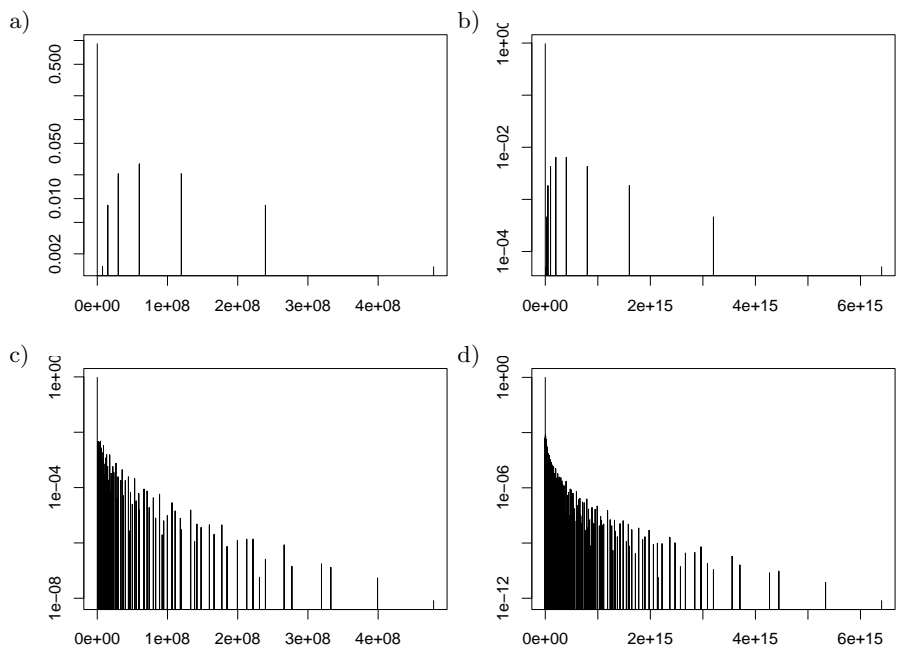}}
\caption{\label{Fig:BlockPathsFull} Exact distribution of the number of accessible paths for the block model with 
a) $L=12$ and $m=2$, b) $L=18$ and $m=2$, c) $L=12$ and $m=3$, d) $L=18$ and $m=3$.}
\end{figure}

\begin{figure}
\centerline{\includegraphics[width=\textwidth]{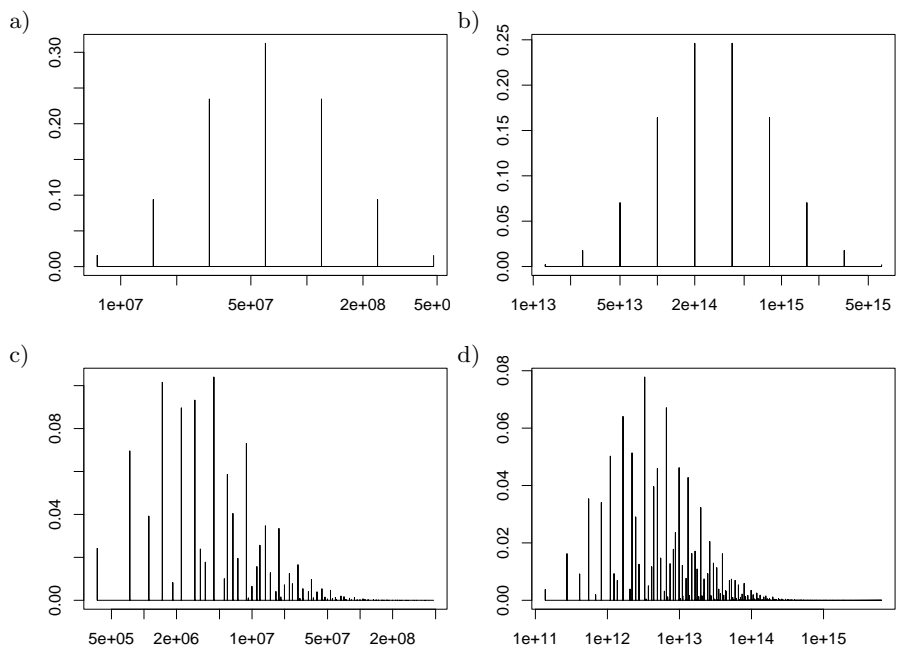}}
\caption{\label{Fig:BlockPathsFullLog} Same as Fig.~\ref{Fig:BlockPathsFull} with the number of paths in 
logarithmic scales, and conditioned on $N_\mathrm{p}>0$.}
\end{figure}

For $L\leq3$ it is feasible to explicitly examine all possible rank orders over the hypercube for their number of accessible paths, and
thus to find the exact path number distributions for the HoC and cHoC models, see Table \ref{Table2}. 
Using these probabilities the exact distribution of the number of accessible paths for the block model 
can be calculated by applying eq.~(\ref{Block_Paths_Full}) for $m\leq3$ and small $L$ (Fig.~\ref{Fig:BlockPathsFull}). 
In particular for $m=2$ the distribution simplifies to
\begin{equation}
\label{Block_Paths_Full_m2}
	\mathbb{P}(N_\mathrm{p}^\mathrm{BN} = N) = \left\{
		\begin{aligned}
			1-\left(\frac{2}{3}\right)^b	& \mbox{  for }	N = 0 \\
			\left(\frac{2}{3}\right)^b\cdot \mathcal{B}_{\frac{1}{2},b}(l) & \mbox{  for  }   l = \log_2{(\frac{m!^b}{L!}\cdot N)} \in \mathbb{N}_0 \\
			0	&	\mbox{  else,}
		\end{aligned}
		\right.
\end{equation}
where $\mathcal{B}_{\frac{1}{2},b}(l)$ is the probability density function of the symmetric binomial distribution with $b$ samples. This means that the logarithm of the scaled number of paths $N_\mathrm{p}/(\frac{L!}{m!^b})$ on \textit{accessible} block landscapes (conditioned
on $N_\mathrm{p} > 0$) with $m=2$ is distributed according to the symmetric binomial distribution 
[Fig.~\ref{Fig:BlockPathsFullLog} a), b)]. For larger $m>2$ the 
distribution becomes more complex and more difficult to write down explicitly, however for $m=3$ the distribution of the 
logarithm of number of paths seems again to be similar to a symmetric, single-peaked distribution 
[Fig.~\ref{Fig:BlockPathsFullLog} c), d)]. This indicates that for
block landscapes that do possess at least one accessible path, the
number of paths is roughly log-normally distributed.

\section{Conclusions}
\label{Summary}

We have shown in this paper that imposing a modular 
block structure on the set of genetic loci substantially changes the behavior of fitness landscapes. 
While mean values for the number of optima as well as for the number of accessible paths are similar between
block landscapes and other types of NK landscapes, there is a qualitative difference between the overall structure 
of the distributions of these topographic features. In both cases the distributions show higher
variability for large $L$ in the block model than in the AN and RN models and also display strong discreteness effects.

The most pronounced difference is observed in the overall evolutionary accessibility, defined here as the probability
for the existence of at least one accessible path to the global
fitness maximum, which decreases very fast with $L$ on block landscapes. 
Together with the rapid increase of the expected number of accessible pathways this implies that, while 
in most instances there is no path to the global maximum, \textit{if} the landscape is accessible 
there are many possible paths. On such untypical landscapes the global
maximum is then relatively likely to be the end result of
the evolutionary process, but the pathway itself is hard to
reconstruct.

Although we used a specific model of modular fitness landscapes our main results hold qualitatively for
 a broader variety of landscapes with modules of independent sets of loci. More precisely, the values of the block fitness functions $f_i$ in (\ref{Fblock}) may be chosen in any way rather 
than being independent and identically distributed random variables, as 
long as all functions are constructed independently from the same ensemble. Also the operation 
connecting the $f_i$ may be any operation that is monotonic in both operands instead of summation (e.g., multiplication).
 Under these broader conditions the number of accessible paths will still be the product of the accessible paths on the modules and basic results such as the exponential decrease in
 accessibility for constant block sizes will still hold. This way it would also be possible to apply our results to modular fitness landscapes that incorporate other biologically 
 important properties, such as neutral mutations.
 
The strict conditions of the SSWM regime may also be lifted. As long as the maximal allowed number of mutations present in the population at any time is limited to a value below
the size of blocks it will be impossible for the population to skip over an entire module and thus any block will still have to be crossable on its own. The number of accessible paths
is then still the product of accessible paths on the single blocks.

Our results suggest that
the choice of neighborhoods in the NK model and, more generally, the architecture of genetic interactions 
is an important aspect to consider when relating fitness landscape models to real world data \cite{Franke2011,Szendro2013a,Neidhart2013}.
Assuming that the genetic architecture itself is, in some sense, under evolutionary selection, 
the low accessibility of modular landscapes would seem to favor connected genetic interaction networks, as unconnected block structures make it impossible to reach the global optimum 
in the SSWM regime. On the other hand, we have also seen that the rare realizations that contain at least
one path tend to have many paths. If each module could evolve independently towards high accessibility, block landscapes would therefore prove advantageous by
allowing many routes to the optimal genotype. 
Interestingly, in the presence of 
recombination the modular structure appears to facilitate rather than impede evolutionary adaptation \cite{Watson2010}, and to elucidate the interplay of recombination
and genetic architecture is a promising direction for future research. 

We can make use of the findings of the present paper to revisit 
the observation, first reported in \cite{Franke2012}, that RN model landscapes are rather 
inaccessible for small values of $k$, in particular for $k=1$ (see Fig.~\ref{Fig:CompAccess}). 
This is surprising because ruggedness is generally expected to increase with 
$k$, such that $k=1$ landscapes should be quite smooth. However, at low $k$ the random graph of interactions between loci is sparse (compare to Fig.~\ref{Fig:NK}), 
and the likelihood for the graph being disconnected, thus effectively giving rise to a modular landscape of low accessibility, is increased.
Inspection of individual instances of the RN model indeed indicates a negative correlation between the accessibility and the number of components of the interaction graph.
However, comparison with the AN model, which by construction has a connected interaction graph 
but displays even lower accessibility than the RN model (Fig.~\ref{Fig:CompAccess}), 
shows that graph connectivity cannot be the main factor determining the accessibility of 
these landscapes. Further investigations are therefore needed to clarify the mechanisms governing   
evolutionary accessibility in generic versions of the NK model.

\paragraph{Acknowledgments.} We acknowledge useful discussions with Peter Hegarty, Anders Martinsson, Johannes Neidhart, Stefan
Nowak and Ivan Szendro, and support by DFG within SFB 680 and SPP
1590. JK takes this opportunity to thank Herbert Spohn for many years of guidance, encouragement and 
inspiration.

\end{document}